\documentclass{aastex631}

\begin{document}

\title{Full velocities and propagation directions of coronal mass ejections inferred from simultaneous full-disk imaging and Sun-as-a-star spectroscopic observations}

\correspondingauthor{Hui Tian}
\email{huitian@pku.edu.cn}

\author{Hong-peng Lu}
\affiliation{School of Earth and Space Sciences, Peking University, Beijing, 100871, People's Republic of China}

\author{Hui Tian}
\affiliation{School of Earth and Space Sciences, Peking University, Beijing, 100871, People's Republic of China}
\affiliation{Key Laboratory of Solar Activity and Space Weather, National Space Science Center, Chinese Academy of Sciences, Beijing 100190, People's Republic of China}

\author{He-chao Chen}
\affiliation{School of Physics and Astronomy, Yunnan University, Kunming 650500, People's Republic of China}
\affiliation{School of Earth and Space Sciences, Peking University, Beijing, 100871, People's Republic of China}

\author{Yu Xu}
\affiliation{School of Earth and Space Sciences, Peking University, Beijing, 100871, People's Republic of China}
\affiliation{Leibniz Institute for Astrophysics Potsdam, An der Sternwarte 16, D-14482 Potsdam, Germany}

\author{Zhen-yong Hou}
\affiliation{School of Earth and Space Sciences, Peking University, Beijing, 100871, People's Republic of China}

\author{Xian-yong Bai}
\affiliation{National Astronomical Observatories, Chinese Academy of Sciences, Beijing 100101, People's Republic of China}

\author{Guang-yu Tan}
\affiliation{School of Earth and Space Sciences, Peking University, Beijing, 100871, People's Republic of China}

\author{Zi-hao Yang}
\affiliation{School of Earth and Space Sciences, Peking University, Beijing, 100871, People's Republic of China}

\author{Jie Ren}
\affiliation{School of Earth and Space Sciences, Peking University, Beijing, 100871, People's Republic of China}



\begin{abstract}

Coronal mass ejections (CMEs) are violent ejections of magnetized plasma from the Sun, which can trigger geomagnetic storms, endanger satellite operations and destroy electrical infrastructures on the Earth. After systematically searching Sun-as-a-star spectra observed by the Extreme-ultraviolet Variability Experiment  (EVE) onboard the \emph{Solar Dynamics Observatory} (SDO) from May 2010 to May 2022, we identified eight CMEs associated with flares and filament eruptions by analyzing the blue-wing asymmetry of the O\,{\sc{iii}} 52.58 nm line profiles. Combined with images simultaneously taken by the 30.4 nm channel of the Atmospheric Imaging Assembly onboard SDO, the full velocity and propagation direction for each of the eight CMEs are derived. We find a strong correlation between geomagnetic indices (Kp and Dst) and the angle between the CME propagation direction and the Sun-Earth line, suggesting that Sun-as-a-star spectroscopic observations at EUV wavelengths can potentially help to improve the prediction accuracy of the geoeffectiveness of CMEs. Moreover, an analysis of synthesized long-exposure Sun-as-a-star spectra implies that it is possible to detect CMEs from other stars through blue-wing asymmetries or blueshifts of spectral lines.

\end{abstract}

\keywords{Solar coronal mass ejections (310) --- Stellar coronal mass ejections (1881) --- Solar filament eruptions (1981) --- Spectroscopy (1558) --- Space weather (2037)  }


\section{Introduction} \label{sec:intro}

On 13 March 1989, a large geomagnetic storm overwhelmed the power grid in Quebec, Canada causing widespread blackouts and accompanied worldwide auroras \citep{Allen89EO00409, Boteler2019SW002278}. Observational evidences show that this geomagnetic storm was caused by coronal mass ejections (CMEs) \citep{Allen89EO00409, Boteler2019SW002278}, which are spectacular ejections of a large amount of plasma and embedded magnetic field from the solar atmosphere \citep{2006SSRv..123..251F, 2011LRSP....8....1C}. The interaction of CMEs with the Earth's magnetosphere significantly disturb the geomagnetic field, triggering geomagnetic storms \citep{1985JGR....90..163S, 1991JGR....96.7831G, 2000GeoRL..27..145G}. CME-induced geomagnetic storms can also endanger satellite operations and the safety of astronauts. Hence, it is of vital importance to predict whether a CME will cause a geomagnetic storm and how strong the storm will be.

An accurate prediction relies on precise knowledge of some key parameters such as the CME propagation direction, full speed and embedded magnetic field direction \citep{1975JGR....80.4204B, 2006JGRA..111.5102K}. The propagation direction largely determines the likelyhood of CME collision with the Earth, and thus is critical for the prediction of geoeffectiveness. However, with the commonly used imaging observations from a single viewpoint, it is very difficult to determine the propagation direction. Halo CMEs are thought to travel towards or away from the Earth \citep{1982ApJ...263L.101H}. However, statistical analyses of halo CMEs show that only about 50\% frontside halo CMEs can produce geomagnetic storms with Kp $\geq$ 5, and the distribution of source locations of these geoeffective halo CMEs is asymmetric on the visible solar disk \citep{2000GeoRL..27.3591C, 2002JGRA..107.1340W}. On the other hand, non-halo CMEs may also hit the Earth and induce geomagnetic storms. These studies suggest that parameters inferred solely from imaging observations, e.g., the source locations and angular widths, cannot adequately determine the propagation directions and geoeffectiveness of CMEs \citep{2002JGRA..107.1340W,2000GeoRL..27.3591C, 2000JGR...10518169S, 2007JGRA..11210102Z}. Although a CME propagation direction parameter that quantifies the asymmetry of the projected CME morphology in coronagraph images was found to have a relatively good correlation with the strength of geomagnetic storm \citep{2005ApJ...624..414M}, its application is only limited to very fast halo CMEs.

Spectroscopic observations could provide valuable information about the propagation of CMEs. According to the Doppler effect, the movement of ejected materials in CMEs along the line of sight towards observers should cause blueshifts or blueshifted component in spectral line profiles \citep{2012ApJ...748..106T}. If we combine the plane-of-sky (POS) and line-of-sight (LOS) velocity components measured with imaging and spectroscopic observations, respectively, we should be able to determine the CME propagation direction and full velocity. However, currently almost all coronal spectrometers are slit spectrometers, which usually scan a small region on the Sun with a typical duration of 1 hour. It is very difficult to catch large-scale coronal transients such as CMEs with such type of observations \citep{2012ApJ...748..106T}. Recently, it was found that the Doppler-shifted ejecta could be observed in the Sun-as-a-star spectra during the initial stage of CME propagation (before appearing in coronagraphs) \citep{2022ApJ...931...76X}. During the solar eruption on 2021 October 28, an obvious secondary component appeared in the blue wing of the O\,{\sc{v}} 62.97 nm line. Combing the velocity component inferred from this secondary component with the velocity component derived from imaging observations from another viewpoint, the propagation direction and full velocity of the CME were derived. 

Here, we adopt a similar strategy but use spectral and imaging observations from a single spacecraft, the Solar Dynamics Observatory (SDO) \citep{2012SoPh..275....3P}, to derive the propagation direction and full velocity of eight CMEs. Sun-as-as-star spectra and full-disk images were obtained with the Extreme Ultraviolet Variability Experiment (EVE) \citep{2012SoPh..275..115W} and the Atmospheric Imaging Assembly (AIA) \citep{2012SoPh..275...17L} onboard SDO, respectively. It is found that the propagation direction and geomagnetic indices (Kp and Dst) are strongly correlated, suggesting that the propagation direction inferred through this way can be used to predict the geoeffectiveness of CMEs. We also synthesize long-exposure Sun-as-a-star spectra from EVE observations and demonstrate that it is possible to detect CMEs from other stars through blue-wing asymmetries or blueshifts of spectral lines.

\section{Data and methods} \label{sec:data}

SDO/EVE measures Sun-as-a-star spectral irradiance over the wavelength range 6-106 nm with a sampling wavelength interval of 0.02 nm, which enables detection of Doppler velocities at tens of kilometers per second \citep{2012SoPh..275..115W}. EVE comprises a grazing-incidence spectrograph (MEGS-A, 6-37 nm) and a two-grating, cross-dispersing spectrograph (MEGS-B, 35-105 nm). The MEGS-B has operated for approximately three hours per day during most period of the mission, while MEG-A has been unable to work since 26 May 2014. In this work, EVE level-2B spectra were used to search for CMEs and plot light curves during the associated flares. Each file in EVE level-2B products contains one-day observational data at a cadence of one minute.

The Sun-as-a-star spectra observed by the MEGS-B instrument between May 2010 and May 2022 were used to identify spectral signatures of CMEs. After systematically searching the 12-year Sun-as-a-star spectral database, only 57 samples with full-phase observations and associated with large flares (flare class $\geq$ M3) were found, from which we only detected eight CMEs associated with flares and filament eruptions by analyzing the blue-wing asymmetry of the O\,{\sc{iii}} 52.58 nm line profiles. Since the O\,{\sc{iii}} 52.58 nm line has been found to be sensitive to plasma motions during CMEs \citep{2022ApJ...931...76X} and the formation temperature of the O\,{\sc{iii}} 52.58 nm line (10$^{4.9}$ K) is close to the peak response temperature of AIA 30.4 nm (10$^{4.7}$ K) \citep{2010A&A...521A..21O, 2012SoPh..275...17L}, we focused on examining the changes in O\,{\sc{iii}} 52.58 nm line profiles during solar eruption events (flares or obvious mass ejections in AIA 30.4 nm images). On the visible disk of the Sun, CMEs moving toward the observer should cause blueshifts of spectral lines. Therefore, we searched for CMEs based on the blue-shifted excess of O\,{\sc{iii}} 52.58 nm line profiles. 

\begin{figure}[ht]%
\centering
\includegraphics[width=0.8\textwidth]{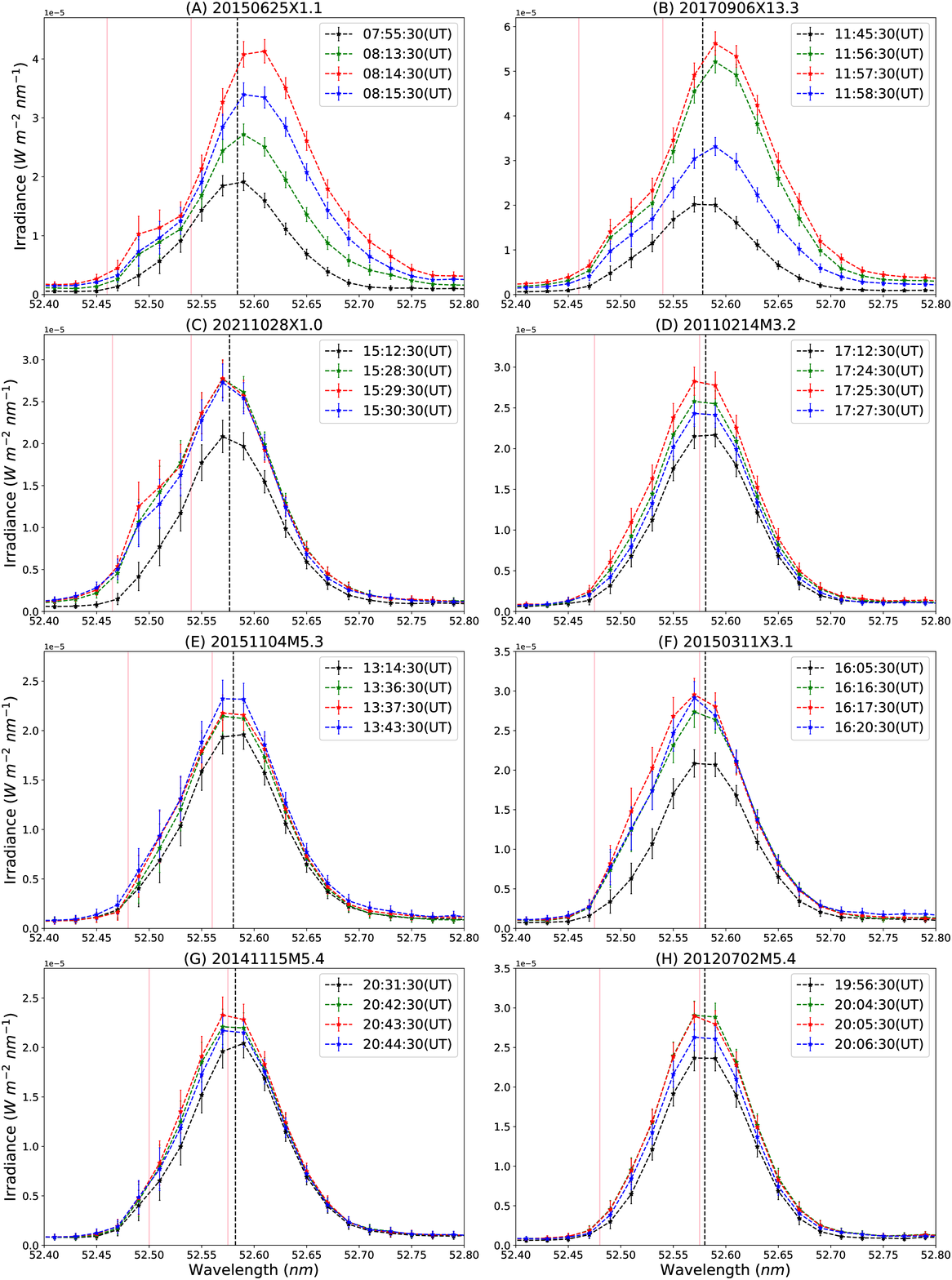}
\caption{\textbf{Examples of Sun-as-a-star O\,{\sc{iii}} 52.58 nm line profiles showing blue-wing asymmetry during eight CME events.} Each event is marked by the associated flare class (GOES class) and date of occurrence, such as 20150625X1.1 for CME 1. The green, red and blue dashed lines with stars in each panel are three line profiles during the CME, and the black dashed line with stars is a pre-eruption line profile (reference spectrum). Each spectral profile is marked with the corresponding observation time. The vertical black dashed line in each panel indicates the reference wavelength. Two pink vertical lines mark the blue wing enhancement.}
\label{fig1}
\end{figure}

Due to the instrumental effect of EVE, the centroid wavelengths of spectral lines may shift in different observation periods \citep{2016SoPh..291.1665C}. Therefore, when examining changes in O\,{\sc{iii}} 52.58 nm line profiles during each eruption event, we selected three O\,{\sc{iii}} 52.58 nm line profiles observed during a relatively quiet period before the onset of each event as reference spectra. To obtain the reference wavelength ($\rm \lambda _{0}$) of the O\,{\sc{iii}} 52.58 nm line for each event, we used the following equation to perform single Gaussian fitting for pre-eruption O\,{\sc{iii}} 52.58 nm line profiles:
  \begin{equation}
      I(\lambda) = A_{0} + A_{1} e^{-\frac{(\lambda - \lambda _{0})^{2}}{2 \sigma _{0}^{2}}} \,.
  \end{equation}
By performing single Gaussian fitting for three profiles, we obtained the centroid wavelengths. By averaging the three centroid wavelengths, the reference wavelength of the O\,{\sc{iii}} 52.58 nm line for each event was obtained. Reference wavelengths for eight events are shown as the black vertical dashed lines in Figure \ref{fig1}, which are in the range of 52.5766 nm to 52.5840 nm.

\section{Data Analysis and Results} \label{sec:results}

\subsection{Eight CMEs detected from Sun-as-a-star EUV spectra}

\begin{figure}[ht]%
\centering
\includegraphics[width=1\textwidth]{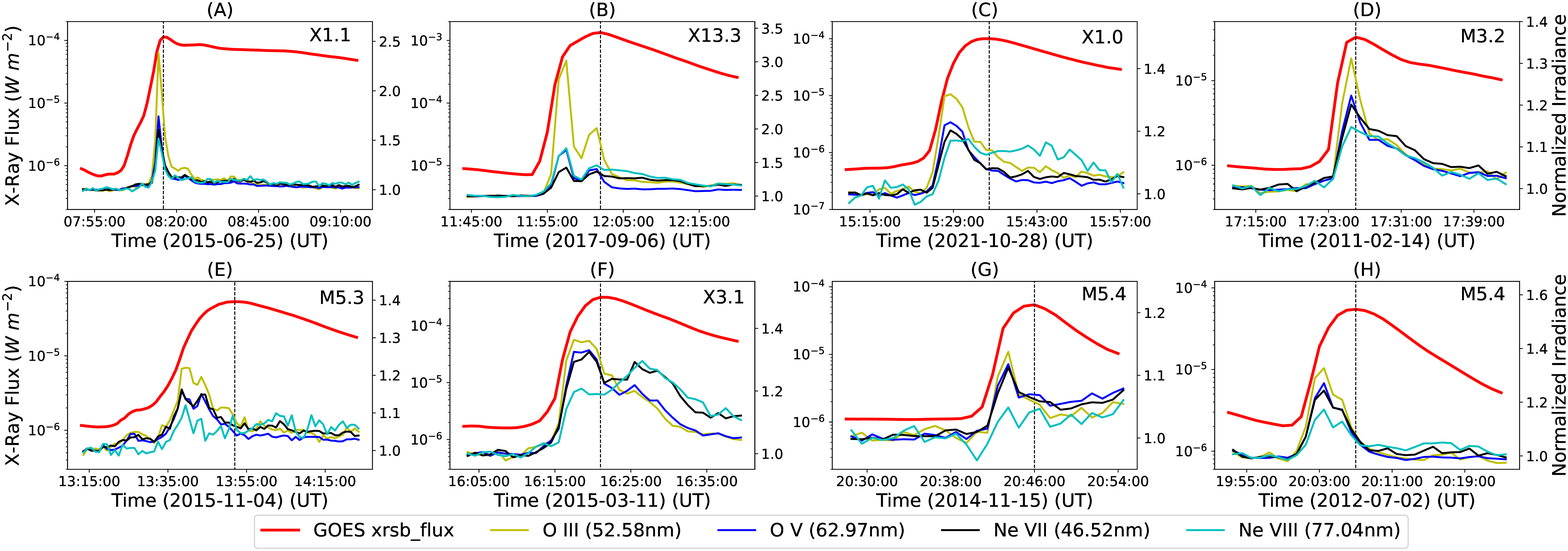}
\caption{\textbf{GOES 0.1-0.8 nm soft X-ray light curves (red) and SDO/EVE O\,{\sc{iii}} 52.58 nm (yellow), O\,{\sc{v}} 62.97 nm (blue), Ne\,{\sc{vii}} 46.52 nm (black) and Ne\,{\sc{viii}} 77.04 nm (cyan) light curves during the eight events.} The black vertical dashed lines indicate the corresponding flare peak times. The top right corner of each panel is labeled with the flare class.}
\label{fig2}
\end{figure}

Using the extreme ultraviolet (EUV) spectra taken by the MEGS-B spectrometer of SDO/EVE between May 2010 and May 2022, we explored changes in O\,{\sc{iii}} 52.58 nm line profiles during solar eruption events and identified spectral signatures of eight CMEs that are associated with solar flares and visible mass ejections on the solar disk. Examples of O\,{\sc{iii}} 52.58 nm line profiles with identified blue-wing enhancement for the eight CMEs are plotted in Figure \ref{fig1}. Figure \ref{fig1} only presents three O\,{\sc{iii}} 52.58 nm line profiles showing blue-wing asymmetry during each CME event and one pre-eruption line profile for comparison. All O\,{\sc{iii}} 52.58 nm line profiles showing blue-wing asymmetry during each CME event can be accessed at \url{https://nadc.china-vo.org/res/r101180/}. Figure \ref{fig2} presents updated soft X-ray light curves measured by the GOES satellite and temporal evolution of the O\,{\sc{iii}} 52.58 nm, O\,{\sc{v}} 62.97 nm, Ne\,{\sc{vii}} 46.52 nm and Ne\,{\sc{viii}} 77.04 nm irradiances observed with SDO/EVE for these eight events. The irradiances are significantly enhanced during the associated flares, and peak before the soft X-ray peak times. It should be mentioned that the blue wing asymmetry of O\,{\sc{iii}} 52.58 nm appears near the peak time of the O\,{\sc{iii}} 52.58 nm irradiance, which means that plasma starts to be ejected during the flare impulsive phase. The relevant parameters of these eight CMEs are listed in Table \ref{tab1}, including the source location (Location), the time of CME's first appearance in the field of view of the Large Angle and Spectrometric Coronagraph (LASCO) \citep{1995SoPh..162..357B} C2 (CME\_C2\_Time) onborad the Solar and Heliospheric Observatory, the central position angle (the middle position angle with respect to the two edges of the CME in the sky plane; CPA) \citep{2004JGRA..109.7105Y}.

\begin{deluxetable*}{cllcccccccc}[b!]
\tablecaption{Information of eight CMEs \label{tab1}}
\tablenum{1}
\tablewidth{0pt}
\tablehead{
\colhead{CME ID} & \colhead{CME Name} & \colhead{Location\tablenotemark{a}} & \colhead{CME\_C2\_Time\tablenotemark{b}} & \colhead{CPA\tablenotemark{c}} & \colhead{$\rm V_{los}$\tablenotemark{d}} & \colhead{$\rm V_{pos}$\tablenotemark{e}} & \colhead{$\rm V_{full}$\tablenotemark{f}} & \colhead{$\rm \alpha$\tablenotemark{g}} & \colhead{Dst\tablenotemark{h}} & \colhead{Kp\tablenotemark{h}}  \\
\colhead{} & \colhead{} & \colhead{} &  \colhead{(UT)}  & \colhead{(deg)}  & \colhead{(km/s)} & \colhead{(km/s)} & \colhead{(km/s)} & \colhead{(deg)} & \colhead{(nT)} & \colhead{} 
} 
\startdata 
 1 & 20150625X1.1    & N13W40 & 08:36 & Halo & -499(31) & 308(7)   & 586(32) &   32(4) &    -46  & 3.7  \\
 2 & 20170906X13.3  & S07W33 & 12:24 & Halo & -458(14)  & 29(2)    & 459(14)  &   4(5)  &    -122 & 8   \\
 3 & 20211028X1.0    & S26W04 & 15:48 & Halo & -459(13)  & 225(54) &  511(56)  & 26(13) &  -36  & 4   \\
 4 & 20110214M3.2   & S20W04 & 18:24 & Halo & -339(35) &  503(31) &  607(47) &  56(7) &    -20 &  2.3   \\
 5 & 20151104M5.3   & N06W04 & 14:48 & Halo & -450(34) &  96(14)   &  460(37) &  12(11) &  -87 &   6  \\
 6 & 20150311X3.1    & S16E26 & 17:00 & 73     & -370(27) &  194(15) &  406(31) &  28(7) &    -15 & 2.3  \\
 7 & 20141115M5.4   & S15E44 & 21:24 & 142    & -339(37) &  273(13) & 435(39) &   39(7) &   -36 &  3.3  \\
 8 & 20120702M5.4   & S17W01 & 20:24 & 185   & -230(44)  &  364(38) & 431(58) & 58(13) &  -13 & 2.7  
\enddata
\tablecomments{(a) The CME source location; (b) the time of CME's first appearance in the field of view of the LASCO/C2; (c) the central position angle; (d) the average line-of-sight velocity; (e) the average plane-of-sky velocity; (f) the full velocity for the bulk motion of the ejected plasma; (g) the angle between CME propagation direction and the Sun-Earth line; (h) the geomagnetic indices Dst and Kp.}
\end{deluxetable*}

\begin{deluxetable*}{cllcccccccc}[b!]                                                                                                                                                                                                                                                                                                                                                                                                                                                                                           
\tablecaption{Gaussian fitting parameters for the O\,{\sc{iii}} 52.58 nm line profiles with blue-wing asymmetry. \label{tab2}}                                                                                                                                                                                                                                                                                                                                                                                                                                                                            
\tablenum{2}                                                                                                                                                                                                                                                                                                                                                                                                                                                                                                                     
\tablewidth{0pt}                                                                                                                                                                                                                                                                                                                                                                                                                                                                                                                                             
\tablehead{                                                                                                                                                                                                                                                                                                                                                                                                                                                                                                                      
\colhead{CME Name} & \colhead{num\_sp\tablenotemark{a}} & \colhead{exp\_time\tablenotemark{b}} & \colhead{$\rm \chi_{r}^{2}$\tablenotemark{c}}  & \colhead{$\rm B_{0}$\tablenotemark{d}} & \colhead{$\rm B_{2}$\tablenotemark{d}} & \colhead{$\rm \lambda _{g}$\tablenotemark{d}} & \colhead{$\rm \sigma_{g}$\tablenotemark{d}} & \colhead{$\rm B_{1}$\tablenotemark{d}} & \colhead{$\rm \lambda_{b}$\tablenotemark{d}} &   \colhead{$\sigma_{b}$\tablenotemark{d}} \\                              
\colhead{} & \colhead{} & \colhead{(UT)} &  \colhead{}  & \colhead{$\times 10^{-7}$ }  & \colhead{$\times 10^{-7}$} & \colhead{$\times 10^{-3}$} & \colhead{$\times 10^{-3}$ } & \colhead{$\times 10^{-7}$ } & \colhead{$\times 10^{-3}$} & \colhead{$\times 10^{-3}$ }}                                                                                                                                                                                                                          
\startdata     
20150625X1.1	&  3	 &      08:13:30  &	 1.27	  &  19(3)  &	248(6)  &	52594(1) &	46(1) &	39(26)   &	52498(4) &	11(9)   \\   
              &      &      08:14:30  &  2.45	  &  32(6)  &	378(11) & 52605(2) &	48(2) &	55(16)   & 	52494(6) &	17(6)    \\  
              &      &      08:15:30  &  1.06	  &  24(4)  &	315(6)  &	52602(1) &	49(1) &	34(19)   & 	52497(5) &	11(9)    \\  
20170906X13.3	&  8	 &      11:55:30  &  0.58   &	 13(1)  & 251(2)  & 52583(1) &	46(1) &	21(79)   &	52499(5) &	9(30) \\
		          &      &      11:56:30  &  1.11   &	 30(3)  & 486(6)  & 52595(1) &	46(1) &	74(11)   &	52497(2) &	14(3) \\
		          &      &      11:57:30  &  1.43   &	 38(4)  & 517(8)  & 52596(1) &	47(1) &	71(15)   &	52497(3) &	14(4) \\
		          &      &      11:58:30  &  0.57   &	 22(2)  & 303(3)  & 52589(1) &	48(1) &	46(8) 	 &  52499(2) &	13(3) \\
		          &      &      11:59:30  &  0.69   &	 24(2)  & 269(4)  & 52586(1) &	48(1) &	52(7) 	 &  52498(2) &	15(3) \\
		          &      &      12:00:30  &  0.77   &	 29(3)  & 324(5)  & 52590(1) &	48(1) &	74(10)   &	52497(2) &	15(3) \\
		          &      &      12:01:30  &  1.03   &	 33(3)  & 338(6)  & 52589(1) &	49(1) &	63(12)   &	52497(3) &	14(4) \\
		          &      &      12:02:30  &  0.65   &	 28(2)  & 262(4)  & 52585(1) &	48(1) &	71(7) 	 &  52498(1) &	15(2) \\              
   ...        & ...  & ...            & ...     &  ...    &  ...     &  ...     &  ...   & ...      & ...       & ...     \\	                                                                                                                                                                                                                                                                                                                                                                                    
\enddata                                                                                                                                                                                                                                                                                                                                                                                                                                                                                                                         
\tablecomments{(a) the number of O\,{\sc{iii}} 52.58 nm line profiles with blue-wing asymmetry during each CME event; (b) the observation time of each spectrum; (c) the reduced chi-square ($\chi_{r}^{2}$) value of Gaussian fitting; (d) Gaussian fitting parameters ($B_{0}$, $B_{2}$, $\lambda _{g}$, $\sigma _{g}$, $B_{1}$, $\lambda _{b}$, $\sigma _{b}$) in Equation \ref{eq2}. The corresponding error is provided in parentheses. The units for $B_{0}$, $B_{1}$, and $B_{2}$ are $W\,$$m^{-2}$$nm^{-1}$, and the units for $\lambda _{g}$, $\lambda _{b}$, $\sigma _{g}$, and $\sigma _{b}$ are $nm$.  The last three parameters ($B_{1}$, $\lambda _{b}$ and $\sigma _{b}$) correspond to the Gaussian component of the blue-wing enhancement. For the single Gaussian fitting of panel C and F in Figure \ref{fig7}, there is no Gaussian component of the blue-wing enhancement, so these three parameters are not available.}
*This is part of the data in Table 2, and the complete data is listed in the spreadsheet.
\end{deluxetable*}   
  
\begin{deluxetable*}{lcccccccc}[b!]
\tablecaption{Information of Emission Measure (EM) calculation \label{tab3}} 
\tablenum{3}
\tablewidth{0pt}
\tablehead{
\colhead{CME Name} & \colhead{AIA\_time\tablenotemark{a}} & \colhead{eruption region\tablenotemark{b}} & \colhead{AIA\_EM\tablenotemark{c}} & \colhead{EVE\_time\tablenotemark{d}} & \colhead{BW\_Irr\tablenotemark{e}} & \colhead{BW\_Rad\tablenotemark{f}} & \colhead{EVE\_EM\tablenotemark{g}}  & \colhead{Ratio\tablenotemark{h}} \\
\colhead{} & \colhead{(UT)} & \colhead{(x1:x2,y1:y2)(arcsec)} & \colhead{($cm^{-5}$)} & \colhead{(UT)} &  \colhead{($erg$ $cm^{-2} s^{-1}$)}  &  \colhead{($erg$ $cm^{-2} s^{-1} sr^{-1}$)}  & \colhead{($cm^{-5}$)}  & \colhead{ }
} 
\startdata 
 20150625X1.1  & 08:13:07 & (630:649,128:242)      & $6.23 \times 10^{27}$  &  08:13:30 & $\rm 1.07 \times 10^{-4}$ & 2116  & $\rm 8.71 \times 10^{27}$    & 1.40   \\                                                                                                                                                                                                                       
 20170906X13.3 & 11:57:29 & (537:585,-259:-220)    & $2.29 \times 10^{28}$  &  11:57:30 & $\rm 2.45 \times 10^{-4}$ & 5591  & $\rm 2.30 \times 10^{28}$    & 1.01   \\                                                                                      
 20211028X1.0  & 15:29:17 & (-135:178,-581:-508)   & $1.04 \times 10^{28}$  &  15:29:30 & $\rm 2.88 \times 10^{-4}$ & 538   & $\rm 2.22 \times 10^{27}$    & 0.21   \\                      
 20110214M3.2  & 17:25:32 & (-20:87,-235:-94)      & $4.64 \times 10^{27}$  &  17:25:30 & $\rm 1.45 \times 10^{-4}$ & 410   & $\rm 1.69 \times 10^{27}$    & 0.36  \\              
 20151104M5.3  & 13:42:30 & (55:171,111:150)       & $1.82 \times 10^{27}$  &  13:42:30 & $\rm 6.55 \times 10^{-5}$ & 619   & $\rm 2.55 \times 10^{27}$    & 1.40   \\      
 20150311X3.1  & 16:22:34 & (-448:-401,-191:-78)   & $2.57 \times 10^{27}$  &  16:22:30 & $\rm 2.92 \times 10^{-4}$ & 2344  & $\rm 9.65 \times 10^{27}$    & 3.75   \\      
 20141115M5.4  & 20:46:34 & (-967:-854,-301:-264)  & $1.29 \times 10^{27}$  &  20:46:30 & $\rm 1.68 \times 10^{-4}$ & 1714  & $\rm 7.06 \times 10^{27}$    & 2.74   \\      
 20120702M5.4  & 20:04:34 & (42:53,-436:-312)      & $5.45 \times 10^{27}$  &  20:04:30 & $\rm 1.48 \times 10^{-4}$ & 4642  & $\rm 1.91 \times 10^{28}$    & 3.50          
 \enddata
\tablecomments{(a) the observation time of AIA 30.4 nm image used to calculate the EM of the erupting material; (b) the eruption region in AIA 30.4 nm image used to calculate the EM; (c) the average EM of the ejected plasma in the AIA 30.4 nm image; (d) The observation time of the spectrum used to calculate the EM of the blue-wing enhancement in O\,{\sc{iii}} 52.58 nm line profile; (e) the irradiance of the blue-wing enhancement in O\,{\sc{iii}} 52.58 nm line profile; (f) the radiance of the blue-wing enhancement in O\,{\sc{iii}} 52.58 nm line profile; (g) the EM of the blue-wing enhancement in O\,{\sc{iii}} 52.58 nm line profile;  (h) the ratio of the EM of the blue-wing enhancement in O\,{\sc{iii}} 52.58 nm line profile to that of the ejected plasma in the AIA 30.4 nm image.} 
\end{deluxetable*}

\begin{figure}[ht]%
\centering
\includegraphics[width=1\textwidth]{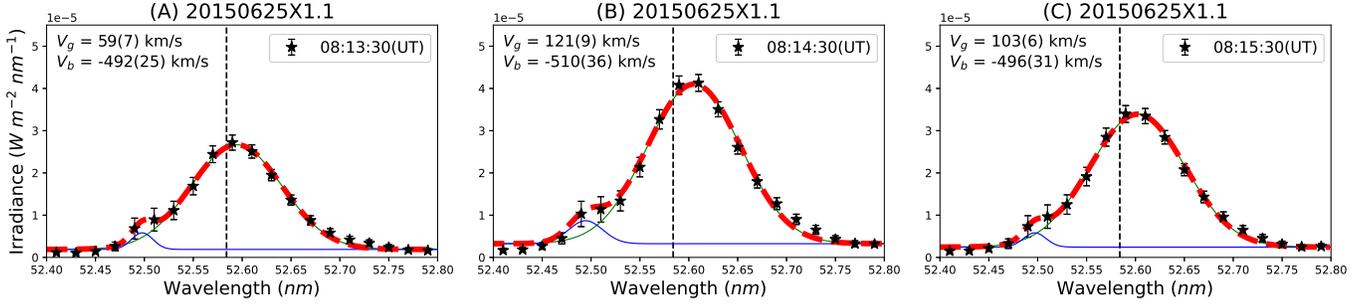}
\caption{\textbf{Double Gaussian fitting results for three O\,{\sc{iii}} 52.58 nm line profiles observed during CME 1 (20150625X1.1).} The black stars represent the observational data of SDO/EVE, and the red dashed lines are double Gaussian fitting results. The blue and green lines represent the two Gaussian components, respectively, and the Doppler velocities corresponding to the centroid wavelengths of the two components are marked in the upper left corner of each panel. The black vertical dashed line indicates the reference wavelength.}\label{fig3}
\end{figure}

For O\,{\sc{iii}} 52.58 nm line profiles with blue-wing asymmetry, the following equation is used for double Gaussian fitting:
  \begin{equation}\label{eq2}
      I(\lambda) = B_{0} + B_{1} e^{-\frac{(\lambda - \lambda _{b})^{2}}{2 \sigma _{b}^{2}}} + B_{2} e^{-\frac{(\lambda - \lambda _{g})^{2}}{2 \sigma _{g}^{2}}}\,.
  \end{equation}
All parameters ($B_{0}$, $B_{1}$, $B_{2}$, $\lambda _{b}$, $\sigma _{b}$, $\lambda _{g}$, $\sigma _{g}$) in Equation \ref{eq2} are considered as free parameters when performing double-Gaussian fitting on the O\,{\sc{iii}} 52.58 nm line profiles. Moreover, the double-Gaussian fitting parameters for all O\,{\sc{iii}} 52.58 nm line profiles with blue-wing asymmetry during the eight CME events are listed in Table \ref{tab2}. The reduced chi-square ($\chi_{r}^{2}$) is used to evaluate the goodness of Gaussian fitting. The closer this value is to 1, the better the fitting result. The reduced chi-square ($\chi_{r}^{2}$) is defined as the following \citep[e.g.,][]{1992drea.book.....B,2001AA...374.1108P,2011ApJ...738...18T}:  
  \begin{equation}\label{eq2a}
       \chi_{r}^{2} = \frac{1}{N-f}\sum_{i=1}^{N}\frac{(d_{i} - m_{i})^{2}}{\sigma_{i}^{2}}\,,
  \end{equation}
where $d_{i}$, $m_{i}$, and $\sigma_{i}$ are the observed spectral irradiance, the fitted spectral irradiance, and the measurement error given by the SDO/EVE, respectively. The $i$ denotes the spectral position, with the summation being carried out across the complete set of $N$ spectral positions. The degree of freedom is represented by $f$ and is equivalent to 4 or 7 for a single- or double-Gaussian fitting, respectively. The $\chi_{r}^{2}$ value of the double Gaussian fitting for each line profile during the eight CME events is also listed in Table \ref{tab2}. It should be noted that, after comparing the single and double Gaussian fitting results of the 63 O\,{\sc{iii}} 52.58 nm line profiles obtained during the eight CME events, we found that, for 53 of them, the $\chi_{r}^{2}$ value of the double Gaussian fitting is closer to 1 than that of the single Gaussian fitting, indicating that double Gaussian fitting is more appropriate for 84\% of these line profiles. For the other 10 profiles, the blue wing enhancement is relatively weak, and the observation times are close to the initial or final stages of filament eruptions. Therefore, for the O\,{\sc{iii}} 52.58 nm line profiles with blue asymmetry during the eight CME events, we adopted the double Gaussian function for fitting. The Doppler velocities corresponding to the centroid wavelengths of the blue and green Gaussian components in Figure \ref{fig3} can be calculated from the following equations:
    \begin{equation}
       V_{b} = \frac{\lambda_{b} - \lambda_{0}}{\lambda_{0}}c\,,
    \end{equation}    
    \begin{equation}
       V_{g} = \frac{\lambda_{g} - \lambda_{0}}{\lambda_{0}}c\,,
    \end{equation}        
where c is the speed of light. The fitting error of the double Gaussian fitting was used to calculate the uncertainties of these two Doppler velocities.

By performing double Gaussian fitting for O\,{\sc{iii}} 52.58 nm line profiles with blue wing asymmetry, we derived the average velocity along the line of sight for the eight CMEs. Taking CME 1 (20150625X1.1) as an example, there are only three O\,{\sc{iii}} 52.58 nm line profiles with blue-wing asymmetry during this CME. Double Gaussian fitting results of the three line profiles and Doppler velocities of the two Gaussian components are shown in Figure \ref{fig3}. By averaging the three Doppler velocities of the blue-wing Gaussian components, we obtained the average velocity of CME 1 along the line of sight as $\rm V_{los}$ = -499 (31) km $\rm s^{-1}$ (the minus sign indicates blueshift; the error is given in parentheses). It should be noted that the reason why we preferred to fit individual time-resolved line profiles is that, for the eight CME events, the signal-to-noise ratio (S/N) of each O\,{\sc{iii}} 52.58 nm line profile is high enough for double Gaussian fitting. In addition, by fitting individual profiles, we can see the evolution of the blue-wing asymmetry of O\,{\sc{iii}} 52.58 nm line profiles. The average velocities of the other CMEs\footnote{Details about the other events can be accessed at \url{https://nadc.china-vo.org/res/r101180/}.} along the line of sight are shown in the column 6 of Table \ref{tab1}.   

\begin{figure}[ht]%
\centering
\includegraphics[width=1\textwidth]{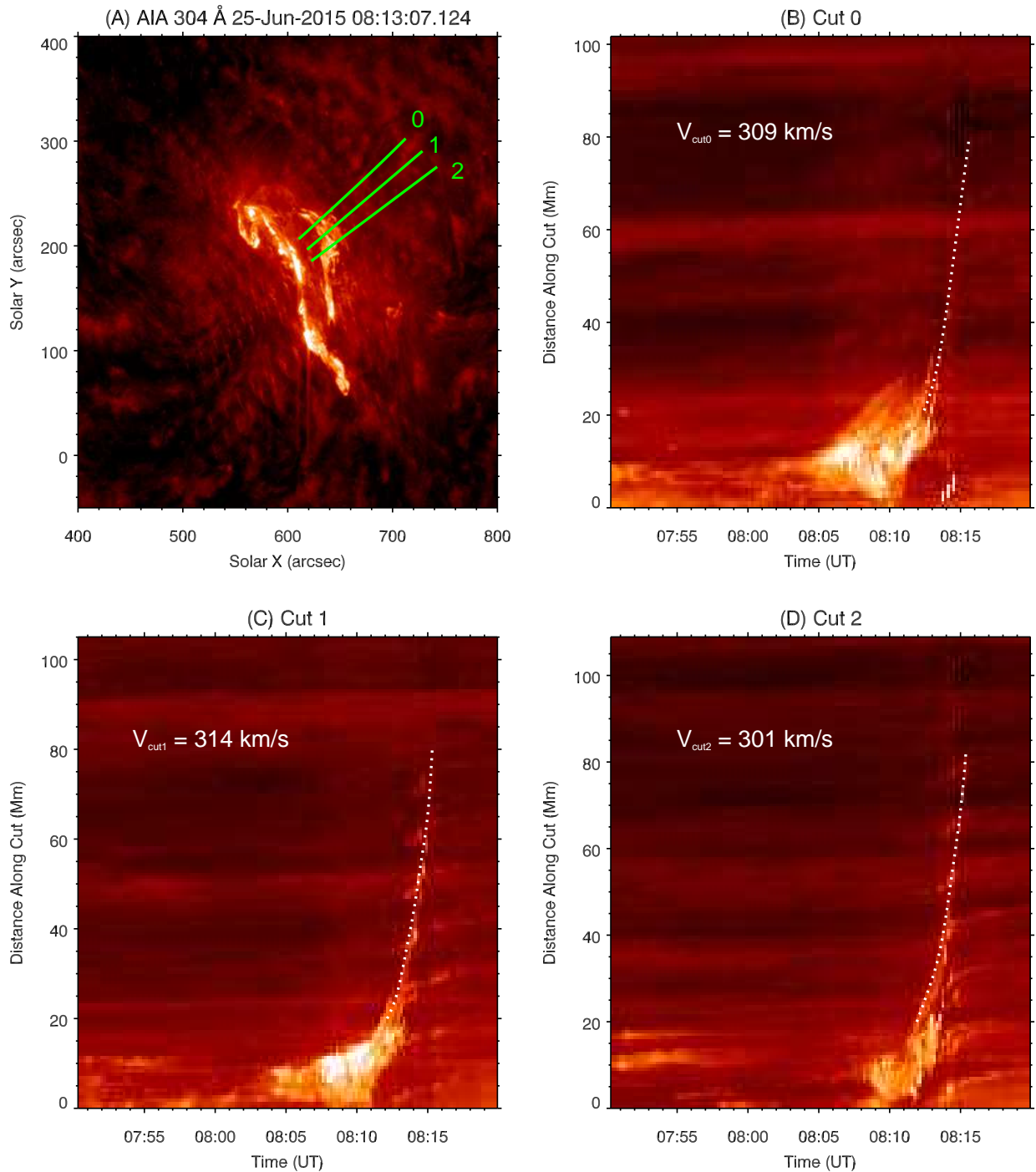}
\caption{\textbf{Ejecta associated with CME 1 (20150625X1.1) in SDO/AIA 30.4 nm images.} (A) AIA 30.4 nm image taken at 08:13 UT, showing the bright ejecta. The green lines represent three cuts along the propagation directions of different parts of the ejecta. (B-D) Time-distance plots for cuts 1, 2, and 3. White dotted lines mark the trajectories of the ejecta along different cuts. The average velocity along each cut is marked in each panel.\\ (An animation of this figure is available.)}\label{fig4}
\end{figure}

In the meantime, the SDO/AIA captured eruptions associated with the eight CMEs through direct imaging at the EUV wavelengths. The 30.4 nm images of SDO/AIA revealed obvious plasma ejections associated with the eight CMEs. The ejecta appear as bright structures, indicating a temperature of $\sim$10$^{4.7}$ K (temperature corresponding to the peak of the response function of AIA 30.4 nm). To calculate the plane-of-sky velocities of the bright ejecta, we chose cuts along the directions of ejecta motion in SDO/AIA 30.4 nm images. Taking CME 1 as an example, we chose three cuts along the direction of the bright ejecta motion, as shown in panel A of Figure \ref{fig4}. The directions of these three cuts are consistent with the propagation direction of CME 1 in LASCO/C2 images. The time-distance plots corresponding to the three cuts are shown in panels B, C and D of Figure \ref{fig4}. The trajectories of ejecta are marked with white dotted lines. Through linear fitting for the trajectories of the ejecta, the average velocities of the ejecta along the three cuts were estimated to be 309 km $\rm s^{-1}$, 314 km $\rm s^{-1}$ and 301 km $\rm s^{-1}$, respectively. By averaging these three velocities, we obtained the plane-of-sky velocity of the ejecta, which is $\rm V_{pos}$ = 308 (7) km $\rm s^{-1}$ in the case of CME 1. The corresponding uncertainty was set as the maximum difference between the three velocities and their average value.The obtained plane-of-sky velocities ($\rm V_{pos}$) of the eight CMEs are listed in column 7 of Table \ref{tab1}.

For each of the eight CMEs, the timing of the blue-wing enhancement in the O\,{\sc{iii}} 52.58 nm line profile coincides with the evolution time of the plasma ejection seen in the AIA 30.4 nm images. The O\,{\sc{iii}} 52.58 nm line has a formation temperature of $\sim$10$^{4.9}$ K, which is close to the peak temperature of the AIA 30.4 nm response function. We assume that the blue wing asymmetries in O\,{\sc{iii}} 52.58 nm line profiles are caused by the ejecta observed in AIA 30.4 nm images. This hypothesis might be evaluated through a comparison between the emission measure (EM) responsible for the blue-wing enhancement in the O\,{\sc{iii}} 52.58 nm line profile and that of the ejected plasma observed in the AIA 30.4 nm passband. Similar to \cite{2016AA...587A..11L}, the EM can be calculated using the following equation:
   \begin{equation}
EM = \frac{I}{E_{f}G(T_{peak})} \quad cm^{-5},
  \end{equation}
where I is the observed intensity, $\rm E_{f}$ is the instrument effective area, $\rm G(T_{peak})$ is the peak of the contribution function of the spectral line. For the AIA 30.4 nm images, the intensity unit of the exposure time normalized image is $DN\,$$\rm pix^{-1}$ $\rm s^{-1}$. The temperature response, $\rm E_{f}G(T_{peak})$, for each CME event can be obtained through the SolarSoftware (SSW) routine aia\_get\_response.pro, and their peak values (unit: $DN\,$$\rm cm^{5} s^{-1} pix^{-1}$) are used for calculation \citep{2012SoPh..275...41B}. For the O\,{\sc{iii}} 52.58 nm line, the EVE level 2B spectral irradiance has been calibrated. The radiance, $\rm I/E_{f}$ (unit: $erg\,$$cm^{-2} s^{-1} sr^{-1}$), of the blue-wing enhancement of the O\,{\sc{iii}} 52.58 nm line profile can be obtained by dividing the corresponding irradiance (unit: $erg\,$$cm^{-2} s^{-1}$) of the blue-wing enhancement by the solid angle ($sr$) occupied by the area of the filament eruption region at 1 au \citep{2017ApJ...844..163S}. This filament eruption region is the same as the eruption region (rectangular region in Table 3) in AIA 30.4 nm image used to calculate the EM.  \cite{2017ApJ...844..163S} obtained the radiance of a spectral line in the SDO/EVE by dividing the irradiance of the line by the solid angle ($6.78 \times 10^{-5} sr$) occupied by the area of the solar disk at 1 astronomical unit (au). Here we need to calculate the solid angle corresponding to the eruption region on the solar disk at 1 au. The calculation process is as follows: first, we need to calculate the area (A) of the eruption region in the AIA 30.4 nm image. Then we can divide this area (A) by the area of the solar disk ($\pi \times 960^{2} \, arcsec^{2}$), and multiply the result by the solid angle ($6.78 \times 10^{-5} sr$) corresponding to the area of the solar disk at 1 au. This will give us the solid angle occupied by the eruption region on the solar disk in the AIA 30.4 nm image at 1 au. The contribution function of the O\,{\sc{iii}} 52.58 nm line can be obtained from CHIANTI \citep{1997AAS..125..149D, 2019ApJS..241...22D}, and we used the peak value $\rm 2.43 \times 10^{-25}$ $erg$ $\rm cm^{3} s^{-1}sr^{-1}$ for our calculation. For the comparison of EMs for each event, we selected a spectrum with obvious blue-wing enhancement of the O\,{\sc{iii}} 52.58 nm line profile, and an almost simultaneously observed AIA 30.4 nm image for EM calculations. For each event, the observation time of AIA 30.4 nm image, the eruption region in AIA 30.4 nm image used to calculate the EM, and the average EM of the ejected plasma in AIA 30.4 nm image are listed in Table \ref{tab3}. The observation time of the EVE spectrum, the irradiance of the blue-wing enhancement of the O\,{\sc{iii}} 52.58 nm line profile, the radiance of the blue-wing enhancement of the O\,{\sc{iii}} 52.58 nm line profile, and the EM of the blue-wing enhancement of the O\,{\sc{iii}} 52.58 nm line profile are also listed in Table \ref{tab3}. The ratio of the EM of the blue-wing enhancement of the O\,{\sc{iii}} 52.58 nm line profile to that of the ejected plasma in the AIA 30.4 nm image is listed in the last column of Table \ref{tab3}. The analysis of the ratio between the two EMs in the last column of Table \ref{tab3} reveals that, for the eight cases, the EM of the blue-wing enhancement of the O\,{\sc{iii}} 52.58 nm line profile and that of the ejected plasma in the AIA 30.4 nm image are comparable. Therefore, we can reasonably assume that the blue-wing asymmetries in O\,{\sc{iii}} 52.58 nm line profiles are caused by the ejecta observed in AIA 30.4 nm images. The discrepancy could be attributed to the following facts: (1) The He\,{\sc{ii}} 30.4 nm line is an optically thick line and not well-modeled by CHIANTI \citep{2005ApJS..157..147W}; (2) The O\,{\sc{iii}} 52.58 nm and He\,{\sc{ii}} 30.4 nm lines have different formation temperatures.

Since the SDO/AIA and SDO/EVE observe the Sun from the same viewpoint, the angle between the line-of-sight velocity and plane-of-sky velocity components is 90 degrees. Therefore, the full velocity ($\rm V_{full} $) for the bulk motion of the ejected plasma can be estimated from the following equation:
   \begin{equation}
V_{full} = \sqrt{V_{los}^{2} + V_{pos}^{2}}
  \end{equation}
In addition, the angle ($\rm \alpha$) between the propagation direction and the Sun-Earth line can be inferred from the following equation: 
   \begin{equation}
tan \alpha = \frac{V_{pos}}{|V_{los}|}
  \end{equation}
After calculation, we found that the full velocity and $\rm \alpha$ angle of CME1 are 586 (32) km $\rm s^{-1}$ and  32 (4) degrees, respectively, during the time period of about 08:13:00 UT - 08:16:00 UT. Errors are calculated through error propagation. The full velocities ($\rm V_{full}$) and the angles ($\rm \alpha$) between the propagation directions of other CMEs and the Sun-Earth line are listed in the eighth and ninth columns of Table \ref{tab1}, respectively. 

\subsection{Strong correlation between CME propagation direction and geomagnetic indices} \label{subsec:correlation}

The propagation direction of CME is an important parameter for predicting geomagnetic storms. Previous studies attempted to examine the correlation between indirect parameters related to CME propagation direction and geomagnetic storms, such as the central position angle and the source location of CME. However, not all CMEs are radially ejected from their source locations, and the correlation between source location and geomagnetic storm strength is weak \citep{2008ApJ...677.1378K}. Since we have obtained the angle ($\rm \alpha$) between the propagation direction and the Sun-Earth line for the eight CMEs, we can explore the correlation between the $\rm \alpha$ and geomagnetic indices.

\begin{figure}[ht]%
\centering
\includegraphics[width=0.8\textwidth]{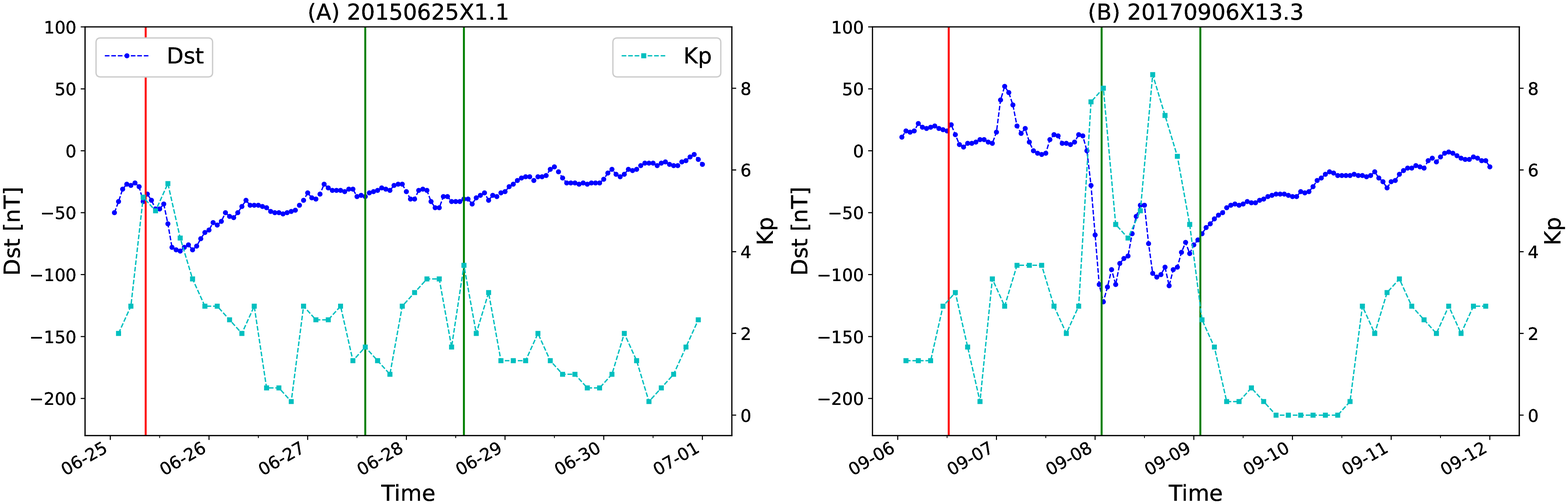}
\includegraphics[width=0.8\textwidth]{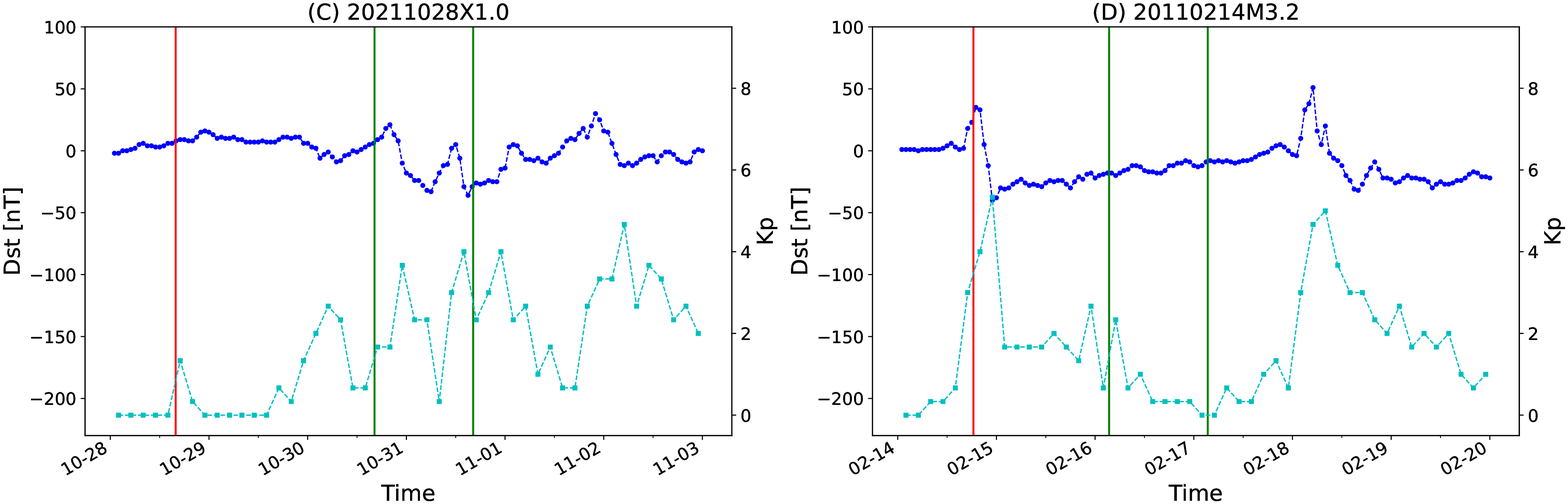}
\includegraphics[width=0.8\textwidth]{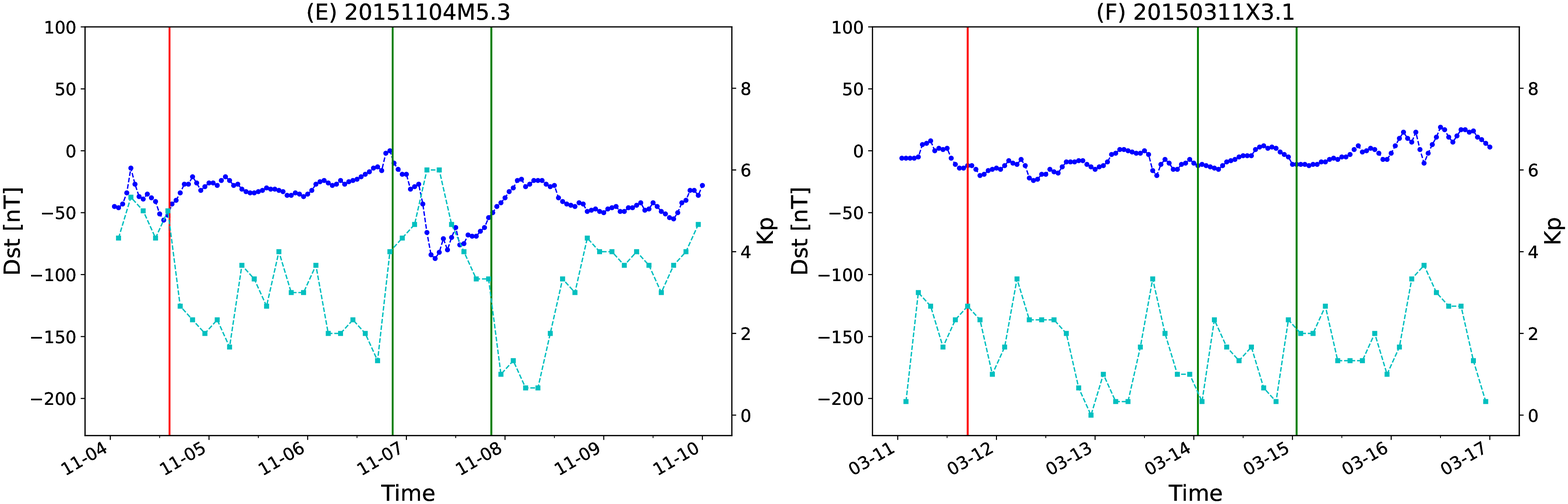}
\includegraphics[width=0.8\textwidth]{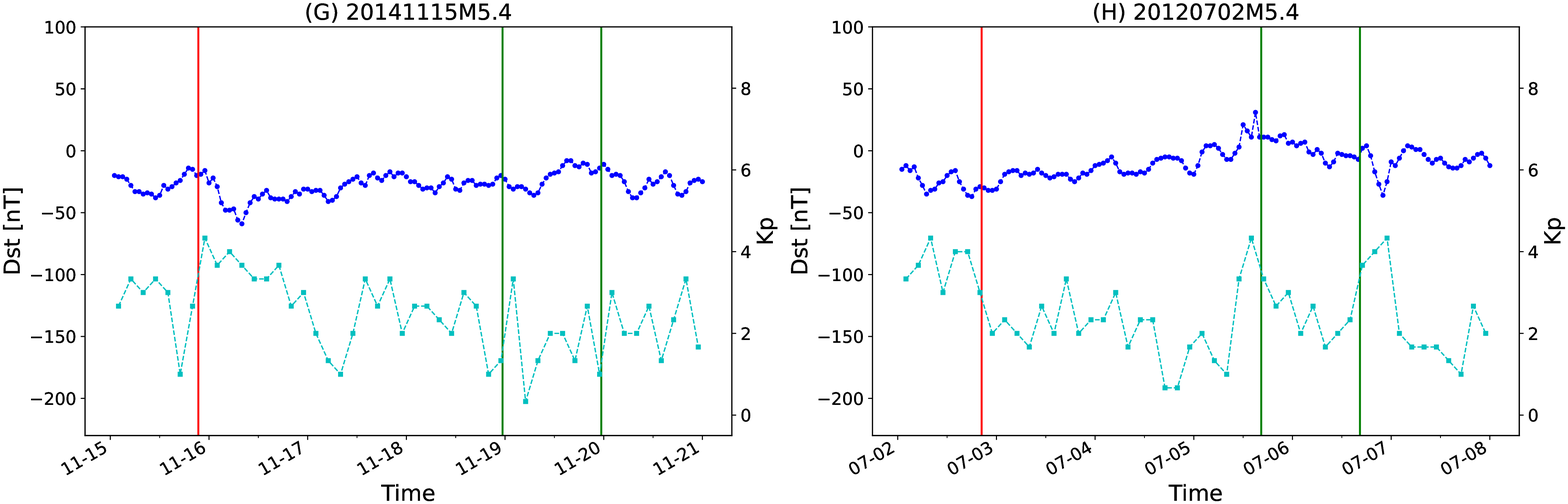}
\caption{\textbf{The temporal evolution of Dst and Kp indices for eight CME events.} The blue circles and cyan squares represents the Dst and Kp values, respectively. The red vertical line marks the time when the CME first appeared in the field of view of the LASCO/C2, and the two green vertical lines indicate the 24-hour time window.}
\label{fig5}
\end{figure}

CMEs are well known to be important drivers of geomagnetic storms. When CMEs reach the Earth's magnetosphere, they could produce nonrecurrent geomagnetic storms if they contain a strong and sustained southward magnetic field component  \citep[e.g.][]{https://doi.org/10.1029/JA093iA04p02511, 1991JGR....96.7831G, https://doi.org/10.1029/2007JA012744}. The indications of geomagnetic storms are shown by the sharp decrease of the Dst index or the sharp increase of Kp index \citep[e.g.][]{1994JGR....99.5771G,Richardson2011}. The geomagnetic indices Kp and Dst were obtained from the Geomagnetic Observatory Niemegk, GFZ German Research Centre for Geosciences (\url{https://www-app3.gfz-potsdam.de/kp_index/Kp_ap_Ap_SN_F107_since_1932.txt}) \citep{matzka2021geomagnetic} and the World Data Center for Geomagnetism in Kyoto (\url{https://wdc.kugi.kyoto-u.ac.jp/dstdir/index.html}), respectively. Figure \ref{fig5} shows the temporal evolution of Dst and Kp indices for the eight CME events. We identified geomagnetic indices associated with the eight CMEs by the following method. For four CMEs (20150625X1.1, 20170906X13.3, 20211028X1.0, 20151104M5.3), the real-time Community Coordinated Modeling Center (CCMC), located at NASA Goddard Space Flight Center, has used the WSA-ENLIL+Cone model \citep{ARGE20041295, ODSTRCIL2003497} to predict the arrival times of the CMEs at the Earth. We obtained the minimum Dst and maximum Kp associated with each of these four CMEs within a 24-hour time window centered at the model predicted arrival time. This time window was set based on the statistical error of the model-predicted CME arrival times, which is slightly less than 12 hours \citep{2001JGR...10629207G, 2018JSWSC...8A..17W}. For the other four CMEs, there are no corresponding models provided by the CCMC. Thus, we estimated the arrival times of CMEs at the Earth through an empirical model \citep{2001JGR...10629207G}, and obtained the minimum Dst and maximum Kp associated with these four CMEs within 24-hour time windows centered at the predicted times. The corresponding minimum Dst and maximum Kp values for the eight CMEs are listed in Table \ref{tab1}.

\begin{figure}[ht]%
\centering
\includegraphics[width=1\textwidth]{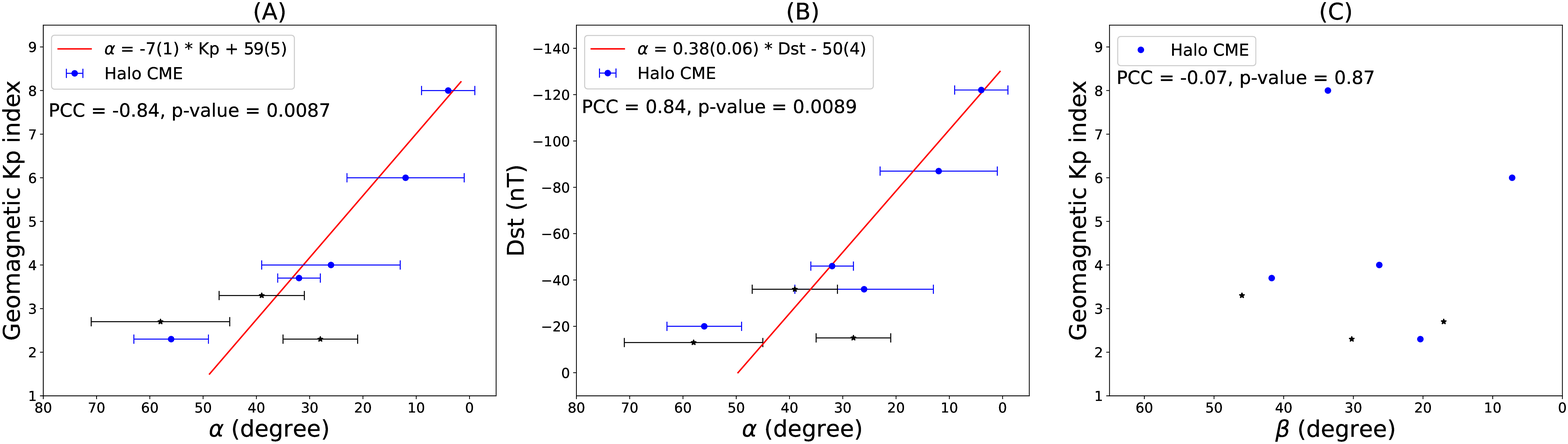}
\caption{\textbf{Correlation between the CME propagation direction and geomagnetic indices.} In panels A and B, $\rm \alpha$ is the angle between the CME propagation direction and the Sun-Earth line. Blue circles and black stars represent halo CMEs and non-halo CMEs, respectively. Red lines are linear fitting results, and the linear fitting equation and Pearson correlation coefficient (PCC) are marked in the upper left corner of each panel. In panel C, $\rm \beta$ is the angle between the Sun-Earth line and the radial direction at the CME source location.}\label{fig6}
\end{figure}

We performed a linear fitting for the relationship between the $\rm \alpha$ angle and the geomagnetic indices for the eight CMEs (Figure \ref{fig6}A-B). The linear fitting equation and Pearson correlation coefficient (PCC) are displayed in Figure \ref{fig6}. Figure \ref{fig6}C shows the angle ($\rm \beta$) between the Sun-Earth line and the radial direction at the CME source location versus the Kp index. The PCC has also been calculated and shown in panel C. 
 
Figure \ref{fig6} shows the correlation between the $\rm \alpha$ and geomagnetic indices. Panel A of Figure \ref{fig6} shows that the $\rm \alpha$ appears to be correlated with the Kp index, the PCC is -0.84 and the corresponding p-value is 0.0087. Panel B shows that the $\rm \alpha$ is also correlated with the disturbance storm time (Dst) index, the PCC is 0.84 and the corresponding p-value is 0.0089. For comparison, panel C of Figure \ref{fig6} shows the angle ($\rm \beta$) between the Sun-Earth line and the radial direction at the source location versus the Kp index. No obvious correlation can be found, the PCC is only -0.07 and the corresponding p-value is 0.87, which is consistent with previous studies showing that the CME source location has only a weak correlation with the strength of geomagnetic storm \citep{2008ApJ...677.1378K}.

\subsection{CME-related blueshifts in long-exposure Sun-as-a-star spectra}

\begin{figure}[ht]%
\centering
\includegraphics[width=1\textwidth]{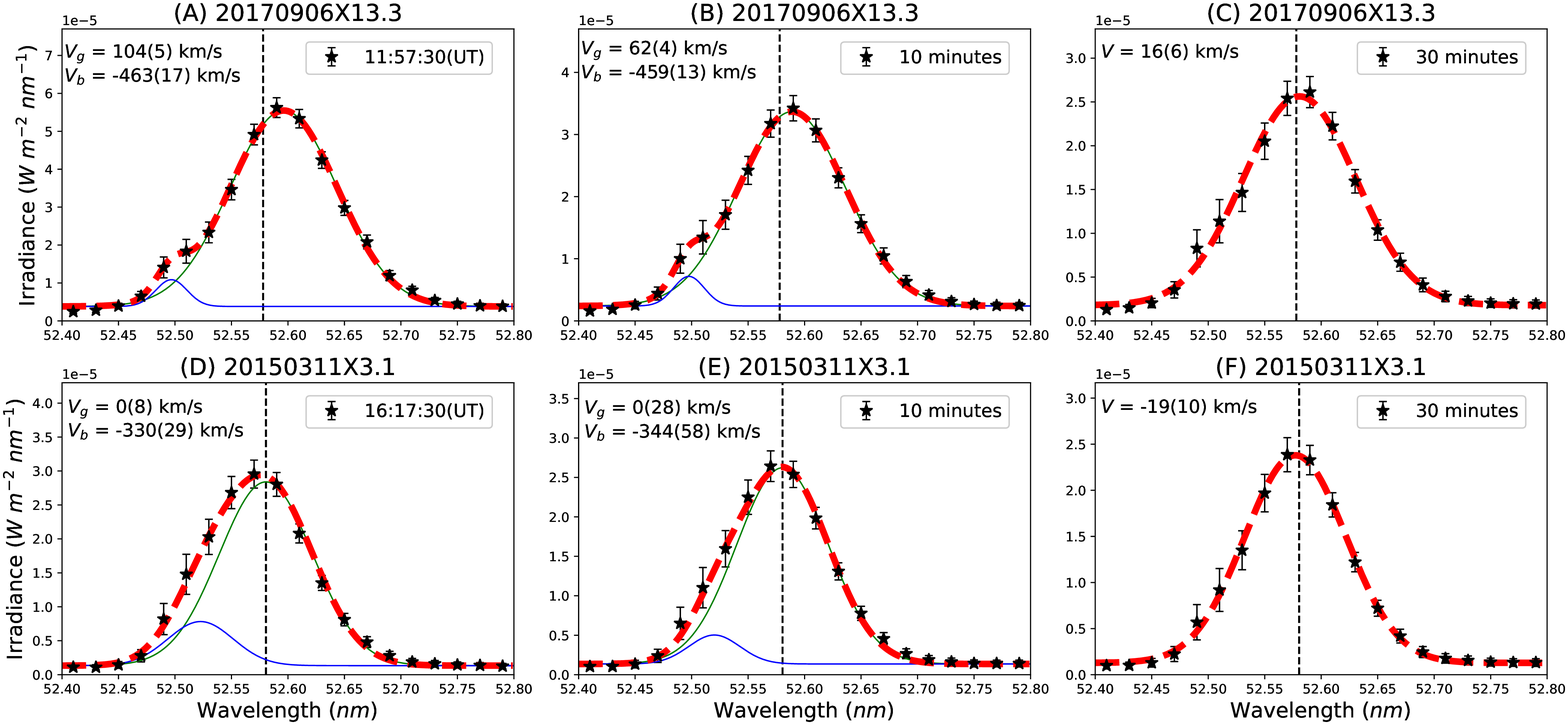}
\caption{\textbf{Gaussian fitting for O\,{\sc{iii}} line profiles with effective exposure times of 1 minute, 10 minutes and 30 minutes.} The black stars represent observational data from SDO/EVE. In panels A, B, D and E, the red dashed lines are double Gaussian fitting results. The blue and green lines represent two Gaussian components, respectively. In panels C and F, the red dashed lines are single Gaussian fitting results. The Doppler velocity corresponding to the centroid wavelength of each Gaussian component is marked in the upper left corner of each panel. The black vertical dashed line indicates the reference wavelength.}\label{fig7}
\end{figure}

Attempts of stellar CME searching have revealed blue-wing asymmetries or blueshifts of spectral lines during flares \citep[e.g.][]{2019NatAs...3..742A, 2019A&A...623A..49V, 2022NatAs...6..241N, 2022ApJ...933...92C, 2022A&A...663A.140L}. Due to the weak emission from other distant late-type stars, stellar optical, ultraviolet and X-ray spectral observations often require long exposures. To examine whether Doppler-shift signals caused by stellar CMEs can be detected in spectral lines observed with long-exposure times, we synthesized long-exposure stellar spectra using the Sun-as-a-star, short-exposure spectra of EVE during CME eruptions. Taking two CMEs (20170906X13.3 and 20150311X3.1) as examples, panels A and D in Figure \ref{fig7} show two short-exposure spectra observed with EVE. For each event, we averaged ten spectra taken during the time period of 11:53:30 UT - 12:02:30 UT and 16:13:30 UT - 16:22:30 UT, respectively, to synthesize spectra with an exposure time of 10 minutes. Also, we averaged 30 spectra taken during the time period of 11:43:30 UT - 12:12:30 UT and 16:03:30 UT - 16:32:30 UT, respectively, to synthesize spectra with an exposure time of 30 minutes. These synthesized long-exposure O\,{\sc{iii}} 52.58 nm line profiles are presented in panels B, C, E and F in Figure \ref{fig7}. Double Gaussian fitting was then applied to the O\,{\sc{iii}} 52.58 nm line profiles for exposure times of 1 minute and 10 minutes (panels A, B, D and E). And single Gaussian fitting was applied to the line profiles observed with an exposure time of 30 minutes. The related parameters of the Gaussian fitting in Figure \ref{fig7} are also listed in Table \ref{tab2}.

Figure \ref{fig7} shows the analysis results of two groups of long-exposure spectra (10 minutes and 30 minutes) synthesized from the short-exposure (1 minute) spectra of EVE during solar CMEs. Panels B and E of Figure \ref{fig7} show that it is possible to detect the CME-caused blue-wing asymmetry from spectra with 10-minute exposure times. Panel C of Figure \ref{fig7} shows a weak redshift of the O\,{\sc{iii}} 52.58 nm line with 30-minute exposure time, which is likely related to the chromospheric condensation during flares \citep{2018PASJ...70..100T, 2022ApJ...928..180W}. Panel F of Figure \ref{fig7} shows that the O\,{\sc{iii}} 52.58 nm line with 30-minute exposure time is slightly blueshifted, which is a result of the associated CME. 

\section{Discussion and Conclusion} \label{sec:Discussion}

In this work, we derived the full CME velocity based on the line-of-sight velocity and plane-of-sky velocity components inferred from single-viewpoint observations. Previous studies have employed various methods to infer the 3D velocities of CMEs \citep[e.g.][]{2012LRSP....9....3W, Zhao_2014, Vourlidas2019}. For instance, with multi-viewpoint coronagraph observations from the Solar Terrestrial Relations Observatory (STEREO; \cite{2008SSRv..136....5K}) and the Solar and Heliospheric Observatory (SOHO), one can reconstruct 3D velocities of CMEs \citep[e.g.][]{2011JASTP..73.1242H, 2012ApJ...751...18F, Colaninno2013, Shi_2015, Wood_2017, 2022AA...660A..23Y}. It is worth noting that this approach requires multiple telescopes to be located at favorable positions. Several studies have also reconstructed 3D velocities of CMEs through single-viewpoint coronagraph observations, which actually rely on geometric assumptions of the shapes of CMEs \citep[e.g.][]{2014ApJ...787..119M, 2016ApJ...824..131R}. Compared to these methods, our method has the following characteristics. First, our method only relies on single-viewpoint observations, which is easier to achieve and more cost efficient compared to multi-viewpoint observations. Second, our method does not rely on assumptions of CME shapes. Third, our method can obtain the 3D CME velocity in the initial propagation phase as a CME travels through the solar disk, while most of the 3D CME velocities reconstructed from coronagraph observations are at later stages than ours. The derived full velocities at the initial propagation phase of CMEs could be used to constrain CME models and help improve the prediction accuracy of the CME arrival time at the Earth, although CMEs may undergo complex acceleration or deceleration during subsequent propagation. For CMEs without multi-viewpoint observations, our method could serve as an alternative approach for reconstructing 3D CME velocities. However, our method has certain limitations. For instance, if the temperature of the erupted plasma is much lower or higher than the formation temperature of the observed spectral line, we would not detect obvious blue-wing enhancements, rendering our method ineffective for deriving the 3D CME velocity. 

We have identified eight CMEs from the blue-wing asymmetries of O\,{\sc{iii}} 52.58 nm line profiles obtained with the MEGS-B spectrometer of SDO/EVE. This is the first catalogue of CMEs identified from Sun-as-a-star EUV spectra using the Doppler shift method. The small number of events in this catalogue is likely related to three facts. First, the MEGS-B has been operating for only approximately three hours per day during most period of the mission. As a result, many CMEs were missed. Second, only relatively strong eruptions have the chance to be detected in the Sun-as-a-star spectra. Blue wing enhancements caused by many relatively small-scale eruptions should be very weak and cannot be detected. Third, to be observed in the O\,{\sc{iii}} 52.58 nm line, at least some of the erupted materials should have a temperature of $\sim$10$^{4.9}$ K. In many eruptions the plasma might be too hot. In addition, it is necessary to determine whether the blue-wing enhancement of the O\,{\sc{iii}} 52.58 nm line profile is caused by chromosphere evaporation during flares. Solar flares observations indicate that blueshifts caused by chromospheric evaporation are often found in high-temperature spectral lines, such as Fe\,{\sc{xii}-\sc{xxiv}} lines (1.25 - 16 MK) and Ca\,{\sc{xix}} line (25 MK), while low-temperature transition lines often do not show blueshifts or blue-wing enhancements \citep[e.g.][]{Doschek_2005, 2009ApJ...699..968M, 2014ApJ...797L..14T, 2015ApJ...811..139T, 2011ApJ...727...98L}. In some cases the transition region lines may reveal a blue shift or blue wing enhancement. However, the velocity is generally 10-30 km/s \citep[e.g.][]{2006ApJ...642L.169M}, much smaller than the velocity we reported here. Thus, we can safely exclude the possibility of chromospheric evaporation.

We have also identified the bright ejecta that are simultaneously observed in the 30.4 nm passband of SDO/AIA. Considering the similar temperature response of the 30.4 nm passband and the O\,{\sc{iii}} 52.58 nm line, we can conclude that the blue-wing asymmetries of O\,{\sc{iii}} 52.58 nm line profiles are caused by the outward moving bright ejecta. It is believed that plasma at typical transition region and chromospheric temperatures within CMEs represents the erupting prominence material \citep{2013JGRA..118..967G}. Thus, at least some of the observed ejecta likely represent the heated filament materials during eruptions, others may simply be the ejected transition region plasma. The typical morphology for CMEs is the so-called three-part structure, including a bright frontal loop, a dark cavity and an embedded bright core \citep{1985JGR....90..275I}. The bright core corresponds to the erupting filament, which often has stronger magnetic field and higher plasma density than the other regions \citep{1981ApJ...244L.117H}. Considering this, the bright ejecta we observed are likely parts that later develop into the cores of CMEs. Thus, the derived full velocity and propagation direction should be those of the CME cores during the early evolution stage of CMEs. Panels A and B of Figure \ref{fig6} show that there is a strong correlation between the propagation direction ($\rm \alpha$) of the CME core and geomagnetic storm strength, suggesting that simultaneous full-disk imaging and Sun-as-a-star spectroscopic observations at EUV wavelengths have the potential for accurate prediction of the geoeffectiveness of CMEs. 

At present, the prediction of geomagnetic storm strength based mostly on coronagraph observations of CMEs often produces false alarms. The real-time Community Coordinated Modeling Center (CCMC) has used models \citep{ODSTRCIL2003497, ARGE20041295, Newell2006JA012015} to predict the geoeffectiveness of four CMEs in our catalogue. It has successfully predicted a moderate geomagnetic storm with Kp of 6 (associated with CME 5) and a weak geomagnetic storm with Kp of 3.7 (CME 1). However, its predictions for CMEs 2 and 3 are largely different from the actual situations. It predicted that CME 2 (20170906X13.3) would produce a geomagnetic storm with maximum Kp up to 6 on the Earth, but a severe geomagnetic storm with maximum Kp of 8 was detected. In addition, the CCMC predicted that CME 3 (20211028X1.0) would produce a severe geomagnetic storm with maximum Kp up to 8 on the Earth, but a weak geomagnetic storm with maximum Kp of 4 was detected. If we use the fitted curve shown in panel A of Figure \ref{fig6}, we would correctly predict the geomagnetic storm strengths for all these four CMEs. Our prediction of Kp is $\sim$8 for CME 2 and $\sim$4.7 for CME 3, in much better agreement with actual observations. Therefore, information about the propagation direction ($\rm \alpha$) of the CME core, which could be obtained from  simultaneous full-disk imaging and Sun-as-a-star spectroscopic observations at EUV wavelengths, can significantly improve the prediction accuracy of geomagnetic storm strength. Considering that we already have routine high-quality EUV imaging observations of the Sun, higher-resolution, higher-sensitivity and routine Sun-as-a-star EUV spectroscopic observations are highly desired to derive the propagation directions of more CMEs and predict their geoeffectiveness. In addition, predicting the arrival times of CMEs is an important component of space weather predictions. Many models have been developed to predict the arrival times of CMEs \citep{Zhao_2014}, including empirical models \citep[e.g.][]{https://doi.org/10.1029/96JA00511, https://doi.org/10.1029/2003JA010300}, expansion speed models \citep[e.g.][]{angeo-23-1033-2005}, drag-based models \citep[e.g.][]{2001SoPh..202..173V, 2010SoPh..261..311S}, physics-based models \citep[e.g.][]{https://doi.org/10.1029/2002GL014865, https://doi.org/10.1029/2011JA017220}, and MHD models \citep[e.g.][]{1995ApJ...447L.143W, 2011ApJ...734...50F}. The derived full velocities at the initial propagation phase could be used as a  key parameter to guide or constrain these models, helping them improve the prediction accuracy of the CME arrival time at the Earth.

We have to mention that the geoeffectiveness of CMEs depends on many factors. For instance, it has been found that large geomagnetic storms are often produced when the embedded magnetic field direction of CMEs is southward. Our result of the correlation between the $\rm \alpha$ angle and the geomagnetic field strength is not in contradiction with this, because a CME has to reach the Earth first before its southward magnetic field reconnects with the Earth magnetic field. Combing information of both the $\rm \alpha$ angle and the embedded magnetic field direction, the prediction for the geoeffectiveness of CMEs will be further improved.
 
Finally, the Sun-as-a-star EUV spectral observations of SDO/EVE during CMEs provide a unique opportunity to explore the detectability of rarely CMEs from other stars. Stellar CMEs are likely the main driver of space weather in the star-exoplanet systems, and could significant impact the habitability of exoplanets \citep{2016NatGe...9..452A}. Our analysis of the synthesized long-exposure Sun-as-a-star spectra has demonstrated that it is possible to detect CMEs through the blue-wing asymmetries or blueshifts of spectral lines in time-domain point-source spectra of distant stars. This result supports previous efforts of stellar CME searching using the Doppler shift method \citep[e.g.][]{2019NatAs...3..742A, 2019A&A...623A..49V,2022NatAs...6..241N, 2022ApJ...933...92C, 2022A&A...663A.140L}, suggesting that the blue-wing asymmetries detected from short-exposure spectra and blueshifts of a few tens km/s in spectral lines obtained during long-exposure observations are indeed very likely to be caused by stellar CMEs.

\begin{acknowledgments}
This research is supported by the NSFC (grants 12250006, 12103004 and 11825301), the fellowship of China National Postdoctoral Program for Innovative Talents (BX2021017) and Frontier Scientific Research Program of Deep Space Exploration Laboratory (2022-QYKYJH-ZYTS-016). Thank you very much for the assistance provided by Dr. Yajie Chen and PhD candidate Xianyu Liu in this work. We thank the data provided by the SDO and GOES teams. SDO is a mission of NASA's Living With a Star Program. This research has made use of the Dst data from the World Data Center for Geomagnetism in Kyoto, and the Kp data from the Geomagnetic Observatory Niemegk, GFZ German Research Centre for Geosciences. The WSA model was developed by N. Arge and the ENLIL model by D. Odstrcil. Estimated real-time planetary Kp indices were obtained from NOAA and archived in the CCMC's iSWA system.
\end{acknowledgments}

\bibliography{lhpref}{}
\bibliographystyle{aasjournal}





\end{document}